# Voltage control of domain walls in magnetic nanowires for energy-efficient neuromorphic devices

*Md. Ali Azam[1], Dhritiman Bhattacharya[1], Damien Querlioz[2], Caroline A. Ross[3] and Jayasimha Atulasimha [1, 4 *]*

[1]Department of Mechanical and Nuclear Engineering, Virginia Commonwealth Univ., Richmond, VA, USA.

[2]Centre national de la recherche scientifique/Univ. of Paris Saclay, Paris, France

[3]Department of Materials Science and Engineering, Massachusetts Inst. of Technology, Cambridge, MA, USA

[4]Department of Electrical and Computer Engineering, Virginia Commonwealth Univ., Richmond, VA, USA.

*email: jatulasimha@vcu.edu

## Abstract

An energy-efficient voltage controlled domain wall device for implementing an artificial neuron and synapse is analyzed using micromagnetic modeling in the presence of room temperature thermal noise. By controlling the domain wall motion utilizing spin transfer or spin orbit torques in association with voltage generated strain control of perpendicular magnetic anisotropy in the presence of Dzyaloshinskii-Moriya interaction, different positions of the domain wall are realized in the free layer of a magnetic tunnel junction to program different synaptic weights. The feasibility of scaling of such devices is assessed in the presence of thermal perturbations that compromise controllability. Additionally, an artificial neuron can be realized by combining this DW device with a CMOS buffer. This provides a possible pathway to realize energy efficient voltage controlled nanomagnetic deep neural networks that can learn in real time.

# I. Introduction

There has been considerable recent progress in the development of dedicated CMOS processors for neuromorphic computing such as IBM's TrueNorth that can implement 1 million spiking neurons and 256 million configurable synapses [1] while consuming ~70mW power. However, these neuromorphic processors have drawbacks such as lack of onboard (real-time) learning/training. More importantly, they have poor energy efficiency in comparison to the human brain, which has ~ 100 billion neurons and ~ 500 trillion synapses and consumes a mere ~ 20 watts of power [2]. Thus, a key challenge for hardware implementation of artificial neural networks lies in finding energy efficient hardware implementations of neurons and non-volatile synapses whose weights can be changed easily and deterministically with very little energy as the network learns from data in real time. While artificial neurons and synapses have been proposed using current-controlled nanomagnets [3-9] and memristors [10-15] that are potentially more energy efficient than pure CMOS implementations, there is still room for increasing the energy efficiency.

We propose implementing energy efficient artificial synapse using a magnetic tunnel junction (MTJ). The free layer of the MTJ comprises a magnetostrictive nanowire racetrack made of CoFe or CoFeB for example. Here we model the magnetization dynamics of the domain walls in the racetrack. The wall is driven through the racetrack clocked with current passing through the racetrack exerting a spin transfer torque (STT) [16-19], or by spin orbit torque (SOT) due to current flowing in a heavy metal layer directly underneath the racetrack [20-23]. The heavy metal layer leads to a perpendicular magnetic anisotropy (PMA) in the CoFeB layer, and a Dzyaloshinskii-Moriya interaction (DMI) which stabilizes chiral domain walls. In order to achieve controlled positioning of the DW we propose current clocked DW motion in conjunction with a gradient in the PMA [24, 25]. Notches are placed at regular intervals to arrest the DW at different locations of the racetrack. For the SOT-clocked DW motion [23, 26-28] instead of using a notched race track and PMA gradient, a racetrack of uniform width is used along with modulation of the PMA at regular intervals. This modulation (PMA reduction) creates a barrier to the motion of the DW, arresting the DW at different locations depending on the voltage applied. We also describe the manner in which this device, in combination with a CMOS buffer, can also function as a neuron and implement Deep Neural Networks (DNNs). Such an implementation is important in applications where energy efficiency is at a premium, such as medical processors and sensor networks that need to learn from data in real time rather than be trained offline, and where synaptic weights of limited accuracy are sufficient [29].

Section II describes the working principle of the device and our micromagnetic modeling approach. Section III presents and discusses simulation of the DW dynamics in the presence of PMA gradients and STT/SOT while Section IV compares the energy efficiency of this approach with other spintronic and memristor approaches.

## II. Device working principle and micromagnetic modeling approach

The working principle of the device (Fig. 1) is explained in terms of the DW dynamics within the magnetic free layer of a magnetic tunnel junction (MTJ). The resistance of the MTJ, which consists of a free layer, a tunnel junction and a fixed layer pinned by a synthetic antiferromagnet (SAF), varies with the location of the DW in the free layer. Therefore, the DW position determines the non-volatile resistance states of the spintronic synapse and can be programed by a voltage, as described below. As the Deep Neural Network (DNN) learns from data in real time, a backpropagation algorithm [30] implemented on a CMOS application-specific co-processor can calculate the new weights for different synapses and output these as specific programming voltages (not addressed in this paper). These voltages should be able to reprogram the resistance states of the synapses to update their resistance values, as described in this work.

**Clocking:** Consider a perpendicular magnetic anisotropy racetrack consisting of a heavy metal/ferromagnet bilayer that could be deposited a piezoelectric film to realize our proposed device as shown in Fig. 1. Such a bilayer (e.g. Pt/CoFe) derives its PMA from interfacial effects and exhibits significant DMI that stabilizes the formation of chiral Néel domain walls [23].

**SOT clock:** SOT acting on the magnetization is generated when current flows in the heavy metal layer. The damping like field (DL-field) thus produced is responsible for translating the Néel domain wall in the ferromagnetic layer [23]. Reversing the direction of the current in the Pt layer reverses the direction of domain wall motion, resets the domain wall position, and hence resets the resistance of the DW MTJ device.

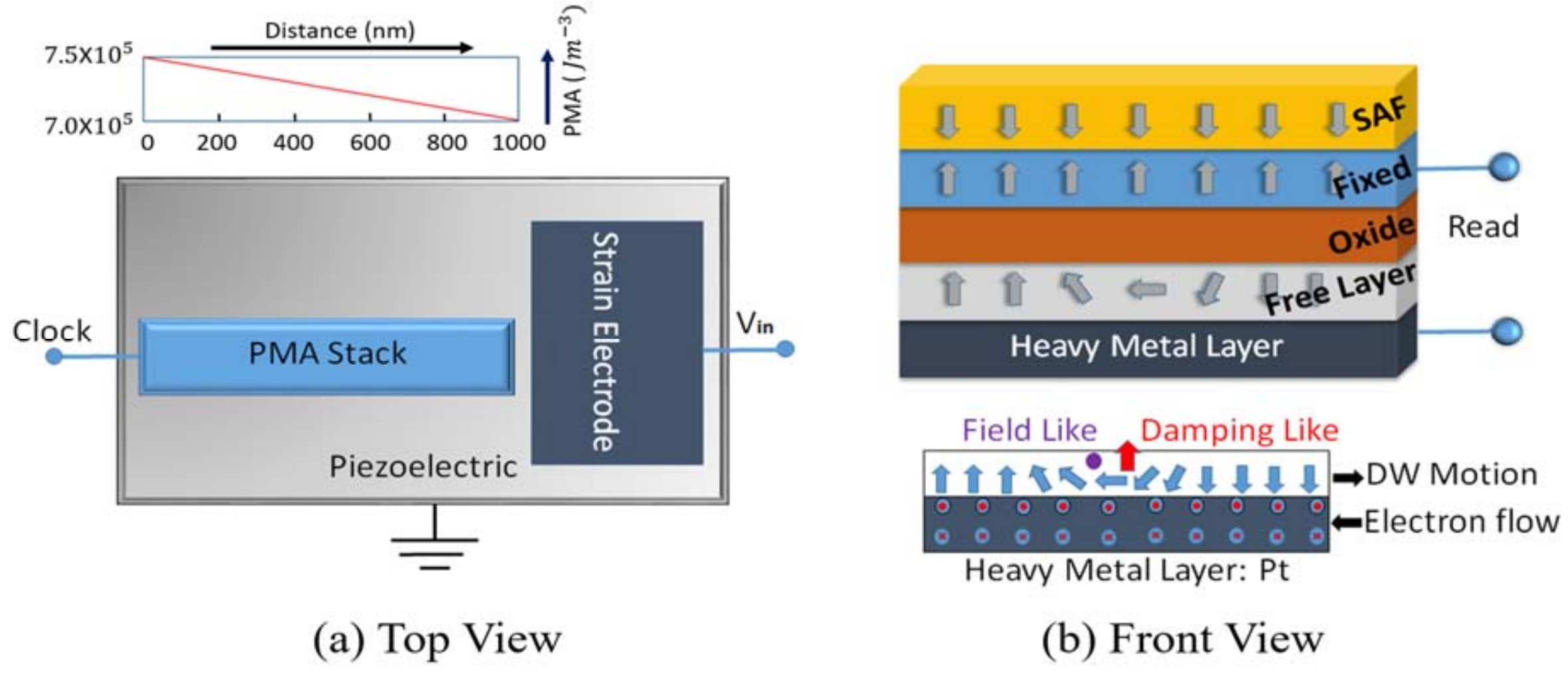

**Fig. 1.** Schematic of the spintronic DW-synapse. (a) Top view showing a magnetic tunnel junction (MTJ) stack placed adjacent to an electrode on a piezoelectric substrate. The inset shows the PMA as a function of distance along the free layer when a voltage is applied to the strain electrode. (b) Front view of the stack (top panel). In the bottom panel, which shows just the heavy metal/free layer, the electrons on the top surface of the Pt are spin polarized into the plane of the figure and electrons at the bottom are spin polarized out of the plane of the figure due to the spin Hall effect.

**STT clock:** Alternatively, for current clocking by spin transfer torque, electrons passing through the domain wall are preferentially polarized along the magnetization orientation of the region through which they pass and exert a torque on the magnetization of the subsequent region they enter [16-19]. This causes the spins within the wall to rotate thus initiating a domain wall motion in the direction of the electron flow.

**SOT vs. STT clock:** When CoFe is used with a heavy metal (Pt) underlayer that leads to PMA and DMI, the ratio of the current flowing through the CoFe that leads to STT and to the current flowing through the Pt that leads to SOT depends inversely on the ratio of their resistances. Rather than consider the case of mixed STT and SOT, we consider the two extreme cases: pure SOT and pure STT in our simulations to understand the domain wall motion with these two clocking mechanisms. We also study the manner in which the DW can be arrested in a specific region of the racetrack by applying a voltage-induced strain under these two different clocking mechanisms.

**Voltage control of domain wall position:** Stopping the domain wall at a specific position along the racetrack is accomplished by applying a voltage to the side electrode (Fig 1(a)) while the DW is being "clocked" by SOT or STT. Consider a domain wall that has been "reset" to one end of the racetrack and is moved along the racetrack towards the other end in Fig. 1 (a) by SOT from a current in the adjacent Pt layer or by STT from a charge current through the free layer. Application of a voltage between the side electrode and the bottom contact of the piezoelectric layer produces an electric field through the piezoelectric thickness, which in turn produces an in-plane stress in the manner described in Ref [31]. This leads to a local strain gradient in the piezoelectric, which is transferred to the ferromagnetic layer, altering its PMA as shown in Fig. 1(a). A modified scheme as shown in Fig 4 can also be used as discussed later.

The mechanism of generation of the strain gradient is explained in detail in Fig. 2. When a voltage is applied to the top electrode, a local electric field is generated through the thickness of the piezoelectric between the area directly underneath the top electrode area and the bottom electrode. This causes an out of plane expansion of the piezoelectric and consequently an in-plane contraction (due to Poisson's ratio) of the area below the top electrode. This produces a tensile in-plane strain in the region of the piezoelectric immediately adjoining the electrode, with a magnitude decreasing with distance away from the electrode. This creates a strain gradient as shown in Fig. 2, upper schematic. While a similar strain gradient is created in the in-plane direction orthogonal to that shown in the figure, we are only concerned with the strain gradient along the DW MTJ device. Furthermore, if the piezoelectric is deposited on a stiff substrate, the bottom of the piezoelectric is clamped but the top part of the piezoelectric can experience the in-plane strain gradient.

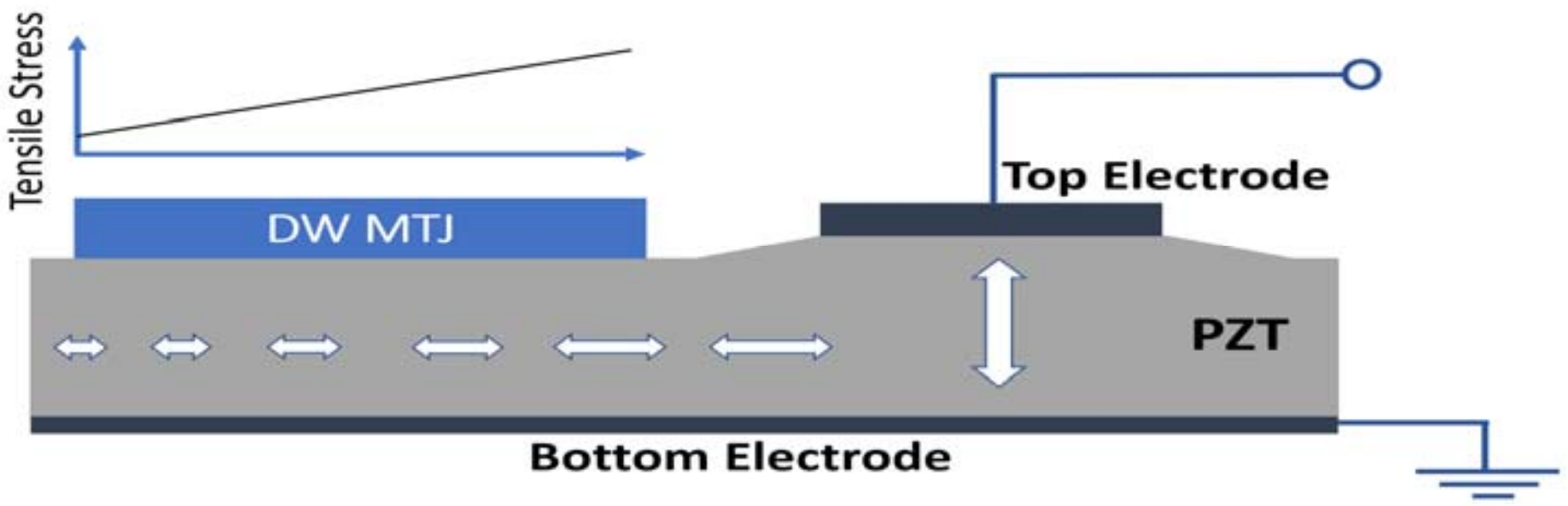

Fig. 2. A schematic showing the strain gradient produced on the DW MTJ when a voltage is applied to the top strain electrode. The in-plane tensile strain decreases from the right end of the DW MTJ to the left producing a PMA gradient.

The benefit of this scheme is that the piezoelectric film develops a strain gradient even though it is not patterned provided the in-plane dimension of the electrode is approximately equal to the thickness of the film [31]. The strain gradient will be most significant within a distance of one to two times the piezoelectric film thickness [31]. This in-plane strain in turn modulates the perpendicular anisotropy of the soft layer and provides a spatial variation of the energy landscape of the Néel DW in the racetrack.

Thus, the device relies on stress generated by the electrode to arrest the SOT/STT-induced motion of the DW, leaving the DW pinned at a notch or a specific location where there is a PMA barrier (Fig 4, discussed later). The strain and therefore modulation of PMA is largest at the left end, and minimum at the right end (Fig 2). This PMA gradient in conjunction with the torque on the DW and notches patterned in the wire determines the position where the DW is arrested.

**Micromagnetic model:** Mumax [32] was used to perform simulations of the domain wall dynamics using the Landau-Lifshitz-Gilbert (LLG) equation in the presence of thermal noise at room temperature. The time rate of change of magnetization in a volume element of the magnetic material is given by:

$$\frac{\partial \vec{m}}{\partial t} = \vec{\tau} = \left(\frac{\gamma}{1+\alpha^2}\right)\left(-\vec{m}\times\vec{H}_{eff} + \alpha\left(\vec{m}\times\left(\vec{m}\times\vec{H}_{eff}\right)\right)\right) \quad (3).$$

where $\vec{m}$ is the reduced magnetization ($\vec{M}/M_{sat}$), $M_{sat}$ is the saturation magnetization, $\gamma$ is the gyromagnetic ratio and $\alpha$ is the Gilbert damping coefficient. The quantity $H_{eff}$ is the effective magnetic field, which is given by:

$$\vec{H}_{eff} = \vec{H}_{demag} + \vec{H}_{exchange} + \vec{H}_{anisotropy} + \vec{H}_{thermal} \quad (4).$$

Here, $H_{demag}$ is the demagnetizing field produced by all the other volume elements of the magnetic material, and $H_{exchange}$ is the effective field due to Heisenberg exchange coupling and DMI [33, 34]. The DMI contribution to the effective exchange field is given by:

$$H_{DM} = \frac{2D}{\mu_0 M_s}\left[\left(\vec{\nabla}.\vec{m}\right)\hat{z} - \vec{\nabla}m_z\right] \quad (5).$$

where $m_z$ is the z-component of magnetization and D is the effective DMI constant. The effective field due to the perpendicular anisotropy is

$$\vec{H}_{anis} = \frac{2K_{u1}}{\mu_0 M_{sat}}(\vec{u}.\vec{m})\vec{u} + \frac{4K_{u2}}{\mu_0 M_{sat}}(\vec{u}.\vec{m})^3\vec{u} \quad (6).$$

where, $K_{u1}$ and $K_{u2}$ are first and second order uniaxial anisotropy constants respectively and $\vec{u}$ is the unit vector in the out-of-plane direction. Strain effectively modulates the anisotropy energy and is incorporated by modulating $K_{u1}$ according to Eq. (2). We assume $K_{u2} = 0$, as we are not dealing with textured or single crystal materials [35].

Thermal noise is modeled by a random, effective magnetic field ($H_{thermal}$) applied in the manner described in [36, 37] within the micromagnetic framework [32]. Furthermore, the field-like and damping-like SOTs and the STT due to charge current are modeled with the appropriate terms [32, 38] assuming polarization=1, Slonczewski parameter Λ=1 and secondary spin-torque parameter έ=0. In our simulations, we do not consider STT and SOT at the same time. Instead, we present two cases: one in which only STT is considered and the other in which only SOT is considered. This leads to an understanding of the clocking of DW motion and its arrest at a specific positions using voltage induced strain for these two different clocking mechanisms.

The discretization cell sizes used for the simulations were 4nm×4nm×1nm and the material parameters used for CoFe (soft layer) of the Pt/CoFe/MgO heterostructure is summarized in Table 1. CoFe has sufficient magnetostriction to produce a PMA gradient that can arrest the DW at specific positions in the model. While Gilbert damping is ~0.01 to 0.03 in these materials, we used a higher value (0.1) so the DW exhibits more stable dynamics. In practice, defects and edge roughness are likely to impede the DW giving dynamics characteristic of the higher damping.

| **Parameters used in simulation** | **Value** |
|---|---|
| Saturation Magnetization ($M_{sat}$) | $1\times 10^{6}$ $Am^{-1}$ |
| Exchange Constant ($A_{ex}$) | $2\times 10^{-11}$ $Jm^{-1}$ |
| Perpendicular Anisotropy Constant ($K_{u1}$) | $7.5\times10^{5}$ $Jm^{-3}$ |
| Gilbert Damping ($\alpha$) | 0.1 |
| DMI Constant (D) | 0.001 $Jm^{-2}$ |
| Saturation Magnetostriction ($\lambda_s$) | 250 p.p.m |

**Table 1.** Material parameters used for the CoFe soft layer in the Pt/CoFe/MgO heterostructure as compiled from previously published works [23, 39-41]. A higher value of Gilbert damping is chosen for the simulation.

## III. Discussion of modeling results

We discuss the modeling results for nanowires of length 1000 nm and width 100 nm patterned with five notches as shown in Fig 3. A gradient in PMA caused by the strain gradient drives the DW towards the lower PMA region in order to reduce the DW energy, i.e. motion is induced in the direction of the negative PMA gradient. We consider the case where the PMA gradient and STT from the charge current drive the DW in the same direction, and then we perform simulation in the presence of thermal noise to understand if these DWs can be arrested deterministically at room temperature at specific notches. We subsequently perform another SOT-driven simulation in which the DW is arrested in regions of varying PMA produced by an alternative electrode design (Fig 4) and no notches are used.

### a. PMA gradient assists STT driven DW motion

In a MTJ racetrack with CoFe soft layer, the DW motion was assumed to be initiated by STT assisted by a PMA gradient (without including the effect of thermal noise). Given that the current pulse acts as the clocking signal, the ON time for current is fixed at 6 ns. This is considered the “write time” for reprogramming the synaptic weight and, based on the analysis of the DW motion at different currents and PMA gradients, it is sufficient to translate the DW to any possible location within this STT clocked device.

Table 2 shows the PMA gradients required to drive the DW from notch 1 to any of the other notches in Fig 3, if the current density of the clock is kept constant at $8.7\ x\ 10^{12}\ A\ m^{-2}$, which is just below the critical current needed to de-pin and initiate the DW motion. For a CoFe layer of 1 nm thickness and width 100 nm this corresponds to a current of ~1 mA (we do not consider any current through the Pt layer and do not consider SOT for this case). We do not account for increase in current density at the notches in this simulation. Furthermore, from the point of view of programming the synapse, the voltage required to achieve a certain DW position need not be a linear function of position, as this is envisaged to be a pre-calculated analog voltage that is output by the co-processor that implements the learning algorithm.

| Final Position of the DW | Required ΔPMA ( $\mathrm{Jm}^{-3}$) over device length | PMA gradient that assists current ($\mathrm{Jm}^{-3}/nm$) |
|---|---|---|
| 2nd Notch | $0.2 \times 10^4$ | 2.00 |
| 3rd Notch | $1.6 \times 10^4$ | 16.00 |
| 4th Notch | $3 \times 10^4$ | 30.00 |
| 5th Notch | $3.7 \times 10^4$ | 37.00 |

**Table 2.** PMA profile for achieving different positions of the DW. The DW starts at the 1st notch. The second column gives the change in PMA over the entire length of the device required to stop the DW at different notches. The third column shows the gradient in PMA for a 1 μm long device.

These results show that at this current density, the current alone cannot drive the DW out of the first notch, but a combination of the current and PMA gradient can. The PMA gradient reduces the current density required to initiate and sustain the DW motion along the racetrack when compared to the case of current only. The DW motion is retarded as it moves into the region of lower PMA, helping to arrest the DW at the notches. The synergistic effects of PMA gradient and current lead to a lower energy operation. Finally, we confirm through simulation (Fig. 3) that each of the different positions of the DW is achievable deterministically, because at a certain PMA gradient the combined strength of the current and PMA gradient is sufficient to move the DW to the desired notch during the current ON period. Therefore, for different strengths of the PMA gradient, the DW travels different distances even though the application time of the current and PMA gradient is kept constant.

**Thermal noise effect at room temperature and scaling issues:** The simulations were repeated in the presence of thermal noise at 300 K. With thermal noise, the minimum current required to initiate the motion of the DW has a lower value when compared to the zero thermal noise case. Moreover, there is a reduction in the effectiveness of the notch in arresting the motion of the DW in presence of thermal noise. Thus, a slight reduction of current is also required to regain the effectiveness of the notches while keeping it high enough so that DW motion can be initiated. This determined our choice of a current density of $8.4 \times 10^{12}\ A\ m^{-2}$ to explore the effect of thermal noise in arresting the DW motion at notch 3 with a PMA gradient of 10 $\mathrm{Jm}^{-3}/nm$. This value of the PMA gradient is chosen between the PMA gradient required for arresting the DW at the 2nd notch ($2\ \mathrm{Jm}^{-3}/nm$) and the 3rd notch ($16\ \mathrm{Jm}^{-3}/nm$). The probability distribution of the final position of the DW is shown in Fig. 3(b). The DW was most likely to be found in notches 3 or 4, but there was a significant probability of its being in notches 2 or 5.

This can be attributed to the relatively small change in energy due to PMA modulation through strain compared to that of thermal energy ($k_bT = 4.14\times10^{-21}$ J). To illustrate the energy scales, considering a 10 $\mathrm{Jm}^{-3}/nm$ PMA gradient (i.e. a total PMA modulation of 10 $\mathrm{kJm}^{-3}$ over the length of 1 μm nanowire), this energy change, $\Delta E = (\Delta PMA/L) \times$ notch spacing $\times$ Volume = $(10\ \mathrm{Jm}^{-3}/\mathrm{nm}) \times 167\ \mathrm{nm} \times 16700\ \mathrm{nm}^3 = 26\times10^{-21}\ \mathrm{J} \approx \sim 6.5\ k_bT$. Here ($\Delta$PMA/L) is the PMA gradient ($\Delta$ PMA over 1 micron length), the volume corresponds to the volume of free layer between two notches, $k_b$ is the Boltzmann constant and T is room temperature in Kelvin. This shows that the change in PMA is modest and hence the PMA gradient does not have deterministic control in positioning the DW in the presence of thermal noise. To circumvent this issue, a higher PMA gradient could be used, though this would require a greater strain gradient, or a greater thickness free layer could be used with its PMA derived from bulk effects (e.g. magnetoelastic) instead of interfacial anisotropy. This analysis has not considered edge roughness [42], which can provide additional pinning sites and even remove the need for lithographic patterning of notches.

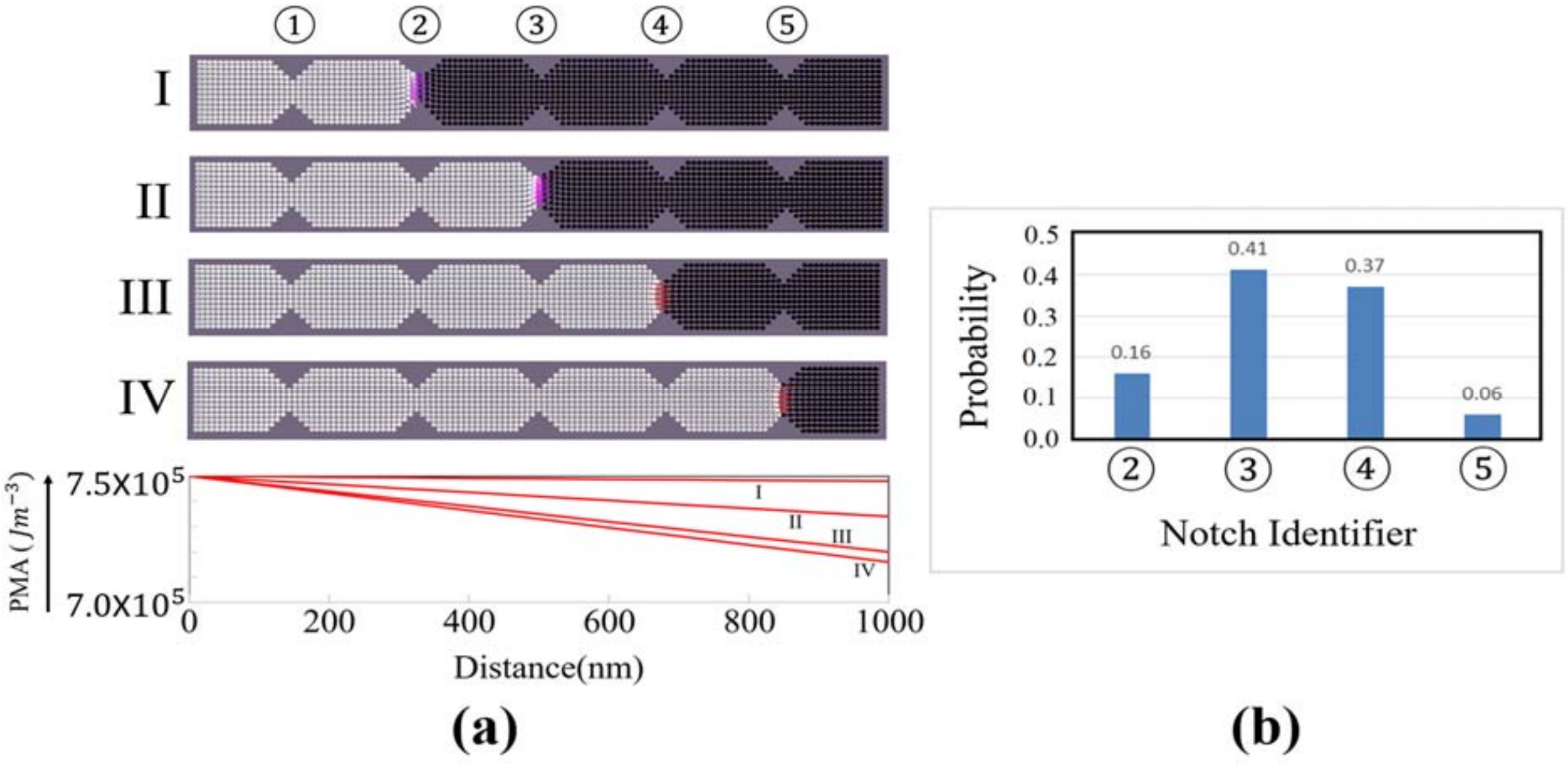

**Fig. 3. (a)** Micromagnetic model showing the final DW positions in the free layer of the MTJ along with the PMA gradients applied for a synergistic PMA gradient and current driven operation of the synaptic weight adjustment. A clear monotonically increasing PMA gradient profile is required to translate the DW from notch ① to notch ②, ③, ④ or ⑤. Both current and DW motion are left to right. NOTE: Supplementary Video 1 a, b, c and d shows a movie of the DW dynamics for each of the above PMA gradients. **(b)** Stochastic behaviour of the DW position due to thermal noise depicted by the widely dispersed distribution of the final DW position. The strain profiles and PMA variations are idealized assumptions and not calculated with detailed finite element analysis. They are estimated based on [31] and scaling arguments discussed later in section IV.

**b. SOT-driven DW motion with pinning of the DW achieved by spatial PMA modulation**

We also simulated DW motion clocked by SOT without including the effect of thermal noise in an MTJ racetrack with CoFe soft layer and no notches. Here we assume no current flows in the CoFe layer and hence there is no STT. We found that discrete PMA variation as shown in Fig. 4 was more suitable to control DW in this case with SOT , instead of using notches and a uniform PMA gradient as in the previous case with STT. The electrode arrangement shown in Fig. 4 alters the PMA at specific regions of the racetrack between the electrodes enabling creation of regions of different PMA. Moreover, due to the different spacing between the pairs of electrodes, the one with smaller gap will create a larger decrease in PMA due to higher stress. However, the PMA profiles in Fig. 4(b) are ideal representation of the actual PMA profile based on scaling arguments presented in section IV and prior work [31]. Determination of the real stress and PMA profile with detailed finite element analysis is beyond the scope of this paper.

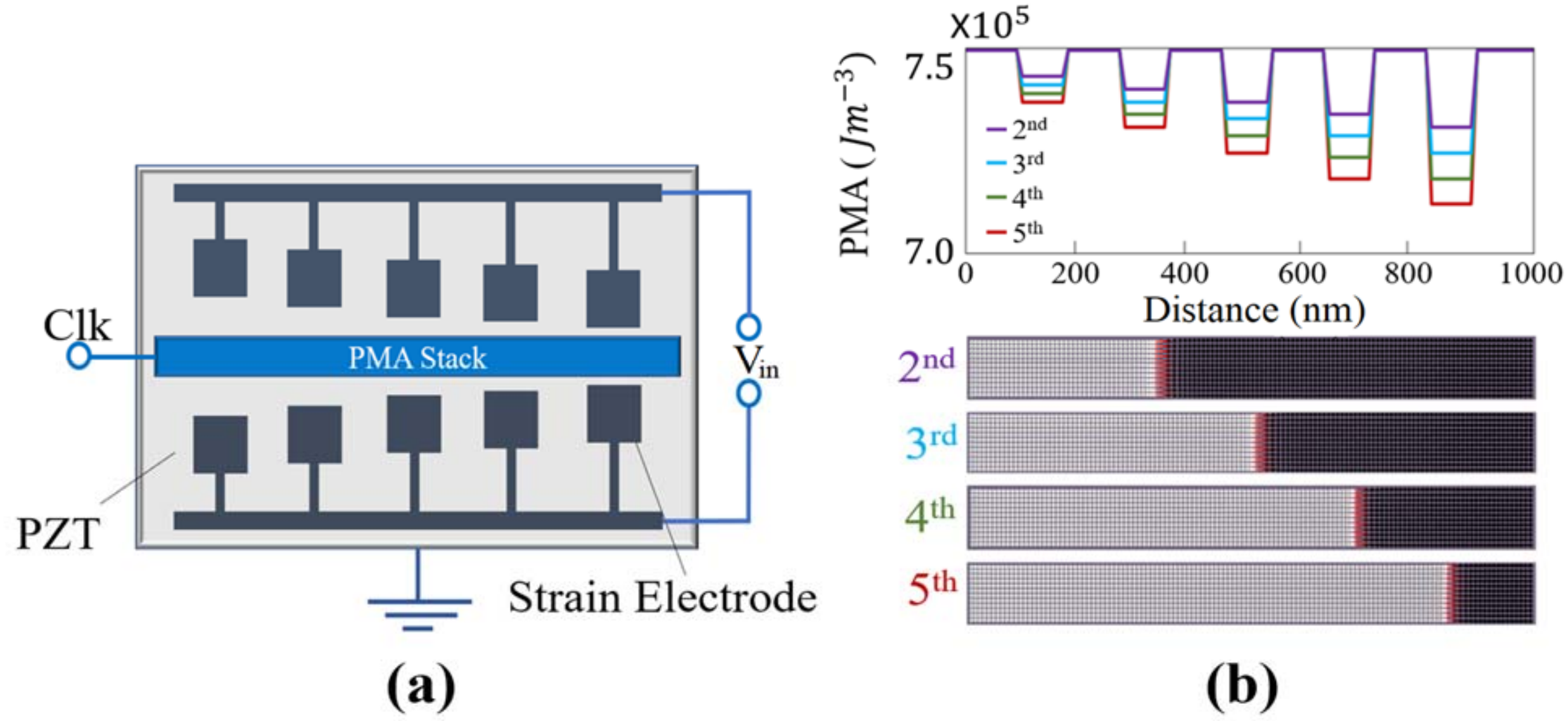

**Fig. 4.** Schematic of the spintronic DW-synapse. (a) Top view showing a magnetic tunnel junction (MTJ) stack placed adjacent to the electrodes on a piezoelectric substrate for SOT driven DW with PMA modulation at regular intervals due to voltage applied to the piezoelectric locally by the electrodes. (b) Different position of the DWs achieved shown in the bottom figure by different PMA profiles created as shown in the upper panel. Initial position of the DW was 180 nm from left (not shown here). The strain profiles and PMA variations are assumptions that were estimated based on [31] and scaling arguments discussed in section IV.

The SOT clock was simulated with a current density $J_c \sim 0.7\times10^{11}$ $Am^{-2}$ through the Pt layer of length 1000 nm, width 100 nm and thickness 1 nm for a clocking period of 20 ns. For lower voltage applied to the strain electrode the PMA reduction is small at the left end. Therefore, the DW experiences smaller barriers and can be translated further to the right. With increasing voltage, barriers at the left increase, thus arresting the DW closer to the left end as shown in Fig 4 (b). The geometric arrangement of the electrodes will ensure that the PMA decrease due to adjacent electrodes varies in a linear fashion. In this case, the ratio between the maximum and minimum PMA change between right most and left most electrode position is kept at 3. This will ensure by only varying the voltage the DW can be arrested at all electrode location without requiring a very large stress to be applied to the piezoelectric. The maximum PMA change for each case of the DW position is shown in Table 3. The DW would eventually drift to the center of the racetrack to minimize magnetostatic energy once the applied voltage is removed. However, with realistic edge roughness the DW is pinned [42] and hence implements a non-volatile synapse.

| Final Position of the DW (Electrode position starting from left) | Maximum ΔPMA ( $Jm^{-3}$) at the rightmost electrode |
|---|---|
| $2^{nd}$ | $0.36 \times 10^5$ |
| $3^{rd}$ | $0.30 \times 10^5$ |
| $4^{th}$ | $0.24 \times 10^5$ |
| $5^{th}$ | $0.18 \times 10^5$ |

**Table 3.** PMA profile for arresting the DW at different electrode position with SOT clocking, showing the PMA decrease at the rightmost electrode. The decrease in the leftmost electrode is 1/3 of this value, and intermediate electrodes change the PMA in a linear fashion as shown in Fig. 4(b)

## IV. Energy Efficiency of Deep Neural Networks (DNNs) with Voltage Control of Domain Walls (VC-DW)

While we have discussed the DW dynamics and operation of the non-volatile voltage programmed synapse in detail, Fig. 5 (a) shows the manner in which this device can be adapted to form a hybrid DW-CMOS neuron. The CMOS buffer implements the threshold functionality of a **neuron** (Fig. 5 a) as well as the ability of the neuron output of one stage to drive inputs to various neurons of the next stage (high fan-out of the CMOS stage) via synapses. In order to reset the neuron, the current/SOT clock is used with current flowing in the opposite direction. Thereafter, the clock is used to synchronize the information flow from one state of the DNN to the next. Fig. 5 describes the manner in which the outputs of one set of neurons can be multiplied by the synaptic weights and input to a neuron at the next stage.

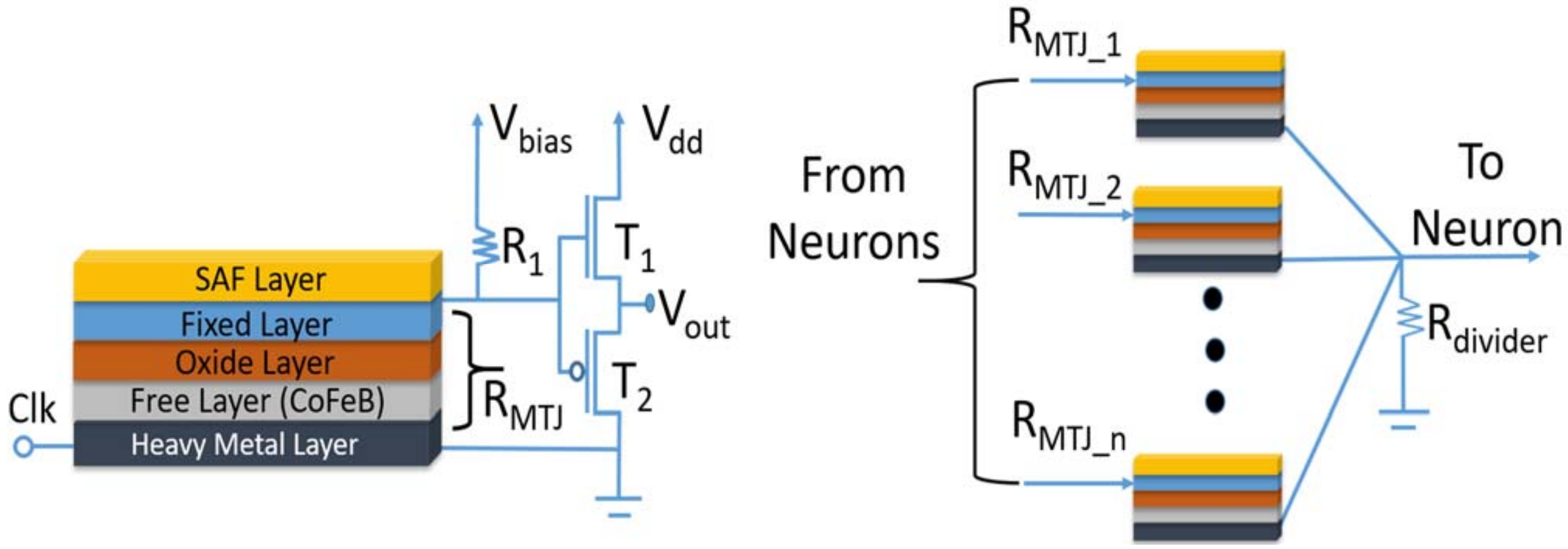

**Fig. 5.** Left: Schematic of the spintronic DW-neuron implemented with SOT. Right: Hybrid neuron and synapse.

**Energy efficiency and area-density of the voltage control DW DNN implementation vs. other implementation schemes**

The energy dissipation in the device can be divided into two parts. One part consists of charging the piezoelectric layer for stress generation, which is essentially the energy lost in charging the capacitor (1/2)$CV^2$, C = capacitance of the piezoelectric layer between the metal contacts, V= voltage applied). The other part is the $I^2R$ loss of the clocking current through the magnetic layer of the racetrack or current through the platinum layer.

For the device simulated in the previous sections, the maximum ΔPMA was $0.37 \times 10^5\ Jm^{-3}$ across the length of the device. The stress required to obtain this may be estimated as $\sigma = \frac{\Delta PMA}{3/2\lambda}$ where λ is the magnetostriction. For CoFeB, with $\lambda \sim 30 \times 10^{-6}$, an unreasonably high stress of ~800 MPa would be required, but for CoFe with $\lambda \sim 250 \times 10^{-6}$ a stress of order ~100 MPa can produce the needed ΔPMA. From the Young's modulus of CoFe of ~ $2.0 \times 10^5$ MPa, a strain of ~$5 \times 10^{-4}$ is required (or more specifically a strain gradient of $5 \times 10^{-4}$ over 1 micron distance). Reference [31] shows that with a 500 µm thick lead zirconate titanate (PZT) substrate and electrodes of side 600 µm, a strain gradient of ~$10^{-3}$ over a distance of 500 µm is feasible with application of 1.5 kV. This suggests that a strain gradient of ~$10^{-3}$ (~$5\times10^{-4}$ if a single electrode is considered) over 1000 nm distance is possible with application of 3 V. The effective capacitance, C= ~26 fF, assuming relative permittivity of 3000. Hence, the energy dissipated is (1/2)$CV^2$ ~117 fJ for the electrical control of a scaled DW nanowire device with a footprint of 1000 nm × 100 nm and electrode of footprint 1000 nm × 1000 nm. In the SOT-clocked case, considering 70 nm square electrodes and 100 nm thick PZT, the total capacitance is C ~18 fF while voltage needed for strain generation is 0.3 V following Ref [31]. This results in an energy dissipation of ~1 fJ.

For the charge current through the CoFe layer with resistivity of 280 Ω.nm [43] and a current density of $J_c \sim 8.7\times10^{12}$ $Am^{-2}$ with a dimension of 1000 nm length, 100 nm width and 1 nm thickness for a clocking period of 6 ns the energy dissipated due to $I^2R$ loss is ~ 15 pJ. For the SOT scheme, considering the resistivity of Pt to be 100 Ω.nm and charge current density $J \sim 0.7\times10^{11}$ $Am^{-2}$ through the Pt layer of length 1000 nm, width 100 nm and thickness 1 nm for a clocking period of 20 ns, the $I^2R$ loss is ~ 100 fJ. Therefore, energy consumption in the device is dominated by the current that produces the SOT. This can be further reduced if low damping materials such as iron garnets are used [44], which have the further advantage of avoiding current shunting through the magnetic layer. The difficulty will lie in optimizing the edge roughness and geometrical design of the racetrack to provide controllability at such low Gilbert damping. In summary, this clocked domain wall device concept provides a pathway to realize novel energy efficient DW neuromorphic devices where reprogramming of synaptic weights can be performed at ~100 fJ per synapse during the learning phase and similarly small ~100 fJ per neuron during the inference phase of the neural network. In fact, during the inference phase, a neuron implemented with only CMOS devices (not a hybrid DW–CMOS device) would only need ~ few fJ per neuron and the synapses would consume no energy as they are non-volatile.

It is interesting to compare these numbers with alternative implementations of artificial neurons and synapses. The most important benefit of our approach for *artificial synapses* is the large reduction in energy consumption. The use of voltage control in conjunction with SOT drastically reduces the energy requirements versus purely spin torque domain wall-based devices [45]. Non-spintronic nanodevices can also provide multilevel synapses, such as oxide-based memristors [46] and phase change memories [29]. Programming such devices requires the physical motion of atoms in order to create or dissolve conductive filaments (oxide-based memristor) or to crystalize amorphous volumes of chalcogenide materials (phase change memory), which has an inherent energy cost, usually higher than picojoules even in highly scaled devices. On the other hand, these alternative technologies may provide more compact synapses than spintronic ones. Our solution therefore offers an extremely energy efficient approach to potentially implement real time learning-capable systems.

In contrast, the benefits of our approach for *implementing artificial neurons* is reduction in area (density). As neurons do not require non-volatility, CMOS-based solutions are typically used for neurons and have comparable energy consumption with ours. On the other hand, they typically require multiple transistors and several micrometre square of area [47].

## V. Conclusion

The feasibility of an energy efficient voltage-controlled DW implementation of an artificial neuron and synapse was demonstrated using micromagnetic simulations. In this approach, modulation of perpendicular anisotropy with stress in combination with SOT or STT is used to program different synaptic weights as well as to mimic a neuron. Scaling this device to smaller dimensions (for example, ~500 nm × 50 nm × 1 nm) could result in much lower energy dissipation as well as high densities for comparable energy dissipation (for implementing neurons) compared to competing approaches. However, to avoid loss of controllability in deterministic positioning of the DW in the presence of thermal noise, careful optimization of material and device geometry are necessary. In summary, this work provides a pathway to the realization of energy efficient voltage controlled artificial neuron networks with real time learning capability and could stimulate more experimental work in this direction.

**Acknowledgements**

J. A., D.B. and M.A acknowledges support from NSF CAREER grant CCF-1253370 (supplement), ECCS 1609303 and Collaborative/Educational Seed Project with Manufacturing Techniques, Inc. (MTEQ) Prime Contract with the US Army C5ISR Night Vision and Electronic Sensors Directorate (NVESD) and Virginia Microelectronics Consortium (VMEC), grant #MTEQ-19-028 TO 54_FP10321. CAR acknowledges support of the National Science Foundation under award 1639921, and the Nanoelectronics Research Corporation (NERC), a subsidiary of the Semiconductor Research Corporation (SRC), through an SRC-NRI Nanoelectronics Research Initiative award, and SMART, an nCORE Center of SRC and NIST.

## References

[1] Merolla P A, Arthur J V., Alvarez-Icaza R, Cassidy A S, Sawada J, Akopyan F, Jackson B L, Imam N, Guo C, Nakamura Y, Brezzo B, Vo I, Esser S K, Appuswamy R, Taba B, Amir A, Flickner M D, Risk W P, Manohar R and Modha D S 2014 A million spiking-neuron integrated circuit with a scalable communication network and interface *Science 80* **345** 668–73

[2] Drachman D A 2005 Do we have brain to spare? *Neurology* **64** 2004–5

[3] Sharad M, Augustine C, Panagopoulos G and Roy K 2012 Spin-based neuron model with domain-wall magnets as synapse *IEEE Trans. Nanotechnol.* **11** 843–53

[4] Chung J, Park J and Ghosh S 2016 Domain Wall Memory based Convolutional Neural Networks for Bit-width Extendability and Energy-Efficiency *Proceedings of the 2016 International Symposium on Low Power Electronics and Design - ISLPED '16* (New York, New York, USA: ACM Press) pp 332–7

[5] Ma X, Zhang Y, Yuan G, Ren A, Li Z, Han J, Hu J and Wang Y 2018 An area and energy efficient design of domain-wall memory-based deep convolutional neural networks using stochastic computing *Proc. - Int. Symp. Qual. Electron. Des. ISQED* **2018-March** 314–21

[6] Sengupta A, Shim Y and Roy K 2016 Proposal for an all-spin artificial neural network: Emulating neural and synaptic functionalities through domain wall motion in ferromagnets *IEEE Trans. Biomed. Circuits Syst.* **10** 1152–60

[7] Chung J, Park J and Ghosh S 2016 Domain Wall Memory based Convolutional Neural Networks for Bit-width Extendability and Energy-Efficiency *Proceedings of the 2016 International Symposium on Low Power Electronics and Design - ISLPED '16* (New York, New York, USA: ACM Press) pp 332–7

[8] Srinivasan G, Sengupta A and Roy K 2016 Magnetic Tunnel Junction Based Long-Term Short-Term Stochastic Synapse for a Spiking Neural Network with On-Chip STDP Learning *Sci. Rep.* **6** 1–13

[9] Sengupta A, Al Azim Z, Fong X and Roy K 2015 Spin-orbit torque induced spike-timing dependent plasticity *Appl. Phys. Lett.* **106**

[10] Querlioz D, Bichler O and Gamrat C 2011 Simulation of a memristor-based spiking neural network immune to device variations *Proceedings of the International Joint Conference on Neural Networks* pp 1775–81

[11] Merkel C, Kudithipudi D, Suri M and Wysocki B 2017 Stochastic CBRAM-Based Neuromorphic Time Series Prediction System *ACM J. Emerg. Technol. Comput. Syst.* **13** 1–14

[12] Yakopcic C, Alom M Z and Taha T M 2016 Memristor crossbar deep network implementation based on a Convolutional neural network *Proc. Int. Jt. Conf. Neural Networks* **2016-Octob** 963–70

[13] Querlioz D, Bichler O, Dollfus P and Gamrat C 2013 Immunity to device variations in a spiking neural network with memristive nanodevices *IEEE Trans. Nanotechnol.* **12** 288–95

[14] Wang Z Q, Xu H Y, Li X H, Yu H, Liu Y C and Zhu X J 2012 Synaptic learning and memory functions achieved using oxygen ion migration/diffusion in an amorphous InGaZnO memristor *Adv. Funct. Mater.* **22** 2759–65

[15] Cheng M, Xia L, Zhu Z, Cai Y, Xie Y, Wang Y and Yang H 2017 TIME: A training-in-memory architecture for memristor-based deep neural networks *Proceedings of the 54th Annual Design Automation Conference 2017 on - DAC '17* (New York, New York, USA: ACM Press) pp 1–6

[16] Yamaguchi A, Ono T, Nasu S, Miyake K, Mibu K and Shinjo T 2004 Real-Space Observation of Current-Driven Domain Wall Motion in Submicron Magnetic Wires *Phys. Rev. Lett.* **92** 077205

[17] Tatara G and Kohno H 2004 Theory of Current-Driven Domain Wall Motion: Spin Transfer versus Momentum Transfer *Phys. Rev. Lett.* **92** 1–4

[18] Li Z and Zhang S 2004 Domain-Wall Dynamics and Spin-Wave Excitations with Spin-Transfer Torques *Phys. Rev. Lett.* **92** 207203

[19] Thiaville A, Nakatani Y, Miltat J and Suzuki Y 2005 Micromagnetic understanding of current-driven domain wall motion in patterned nanowires *Europhys. Lett.* **69** 990–6

[20] Manchon A and Zhang S 2009 Theory of spin torque due to spin-orbit coupling *Phys. Rev. B - Condens. Matter Mater. Phys.* **79** 1–9

[21] Matos-Abiague A and Rodríguez-Suárez R L 2009 Spin-orbit coupling mediated spin torque in a single ferromagnetic layer *Phys. Rev. B - Condens. Matter Mater. Phys.* **80** 1–6

[22] Miron I M, Gaudin G, Auffret S, Rodmacq B, Schuhl A, Pizzini S, Vogel J and Gambardella P 2010 Current-driven spin torque induced by the Rashba effect in a ferromagnetic metal layer *Nat. Mater.* **9** 230–4

[23] Emori S, Bauer U, Ahn S M, Martinez E and Beach G S D 2013 Current-driven dynamics of chiral ferromagnetic domain walls *Nat. Mater.* **12** 611–616

[24] Zhang Y, Luo S, Yang X and Yang C 2017 Spin-orbit-torque-induced magnetic domain wall motion in Ta/CoFe nanowires with sloped perpendicular magnetic anisotropy *Sci. Rep.* **7** 1–10

[25] Brigner W H, Hu X, Hassan N, Bennett C H, Incorvia J A C, Garcia-Sanchez F and Friedman J S 2019 Graded-Anisotropy-Induced Magnetic Domain Wall Drift for an Artificial Spintronic Leaky Integrate-and-Fire Neuron *IEEE J. Explor. Solid-State Comput. Devices Circuits* **5(1)** 19-24

[26] Ryu K-S, Thomas L, Yang S-H and Parkin S 2013 Chiral spin torque at magnetic domain walls *Nat. Nanotechnol.* **8** 527–33

[27] Bhowmik D, Nowakoski M, You L, Lee O, Keating D, Wong M, Bokor J and Salahuddin S 2014 Deterministic Domain Wall Motion Orthogonal To Current Flow Due To Spin Orbit Torque *Sci. Rep.* **5** 11823

[28] Yu G, Upadhyaya P, Wong K L, Jiang W, Alzate J G, Tang J, Amiri P K and Wang K L 2014 Magnetization switching through spin-Hall-effect-induced chiral domain wall propagation *Phys. Rev. B - Condens. Matter Mater. Phys.* **89** 1–6

[29] Ambrogio S, Narayanan P, Tsai H, Shelby R M, Boybat I, Di Nolfo C, Sidler S, Giordano M, Bodini M, Farinha N C P, Killeen B, Cheng C, Jaoudi Y and Burr G W 2018 Equivalent-accuracy accelerated neural-network training using analogue memory *Nature* **558** 60–7

[30] Rumelhart D E, Hinton G E and Williams R J 1986 Learning representations by back-propagating errors *Nature* **323** 533–6

[31] Cui J, Hockel J L, Nordeen P K, Pisani D M, Liang C Y, Carman G P and Lynch C S 2013 A method to control magnetism in individual strain-mediated magnetoelectric islands *Appl. Phys. Lett.* **103** 3–8

[32] Vansteenkiste A, Leliaert J, Dvornik M, Garcia-Sanchez F and Van Waeyenberge B 2014 The design and verification of Mumax3 *AIP Adv.* **4** 107133

[33] Dzyaloshinsky I 1958 A thermodynamic theory of “weak” ferromagnetism of antiferromagnetics *J. Phys. Chem. Solids* **4** 241–55

[34] Moriya T 1960 Anisotropic Superexchange Interaction and Weak Ferromagnetism *Phys. Rev.* **120** 91–98

[35] Chikazumi S 2009 *Physics of Ferromagnetism* Oxford University Press, 2nd edition, 672 pages

[36] Spedalieri F M, Jacob A P, Nikonov D E and Roychowdhury V P 2011 Performance of magnetic quantum cellular automata and limitations due to thermal noise *IEEE Trans. Nanotechnol.* **10** 537–46

[37] Mayergoyz I D, Bertotti G and Serpico C 2009 *Nonlinear magnetization dynamics in nanosystems,* Elsevier

[38] Ralph D C and Stiles M D 2008 Spin transfer torques *J. Magn. Magn. Mater.* **320** 1190–216

[39] Belmeguenai M, Gabor M S, Roussigné Y, Stashkevich A, Chérif S M, Zighem F and Tiusan C 2016 Brillouin light scattering investigation of the thickness dependence of Dzyaloshinskii-Moriya interaction in C o0.5 F e0.5 ultrathin films *Phys. Rev. B* **93** 1–8

[40] Emori S, Martinez E, Lee K J, Lee H W, Bauer U, Ahn S M, Agrawal P, Bono D C and Beach G S D 2014 Spin Hall torque magnetometry of Dzyaloshinskii domain walls *Phys. Rev. B - Condens. Matter Mater. Phys.* **90** 1–13

[41] Bilzer C, Devolder T, Kim J Von, Counil G, Chappert C, Cardoso S and Freitas P P 2006 Study of the dynamic magnetic properties of soft CoFeB films *J. Appl. Phys.* **100**

[42] Dutta S, Siddiqui S A, Currivan-Incorvia J A, Ross C A and Baldo M A 2017 The Spatial Resolution Limit for an Individual Domain Wall in Magnetic Nanowires *Nano Lett.* **17** 5869–74

[43] Zahnd G, Vila L, Pham T V., Marty A, Laczkowski P, Savero Torres W, Beigné C, Vergnaud C, Jamet M and Attané J-P 2016 Comparison of the use of NiFe and CoFe as electrodes for metallic lateral spin valves *Nanotechnology* **27** 035201

[44] Avci C O, Rosenberg E, Caretta L, Büttner F, Mann M, Marcus C, Bono D, Ross C A and Beach G S D 2019 Interface-driven chiral magnetism and current-driven domain walls in insulating magnetic garnets *Nature Nanotechnology* **14** 561–566

[45] Lequeux S, Sampaio J, Cros V, Yakushiji K, Fukushima A, Matsumoto R, Kubota H, Yuasa S and Grollier J 2016 A magnetic synapse: Multilevel spin-torque memristor with perpendicular anisotropy *Sci. Rep.* **6** 1–7

[46] Prezioso M, Merrikh-Bayat F, Hoskins B D, Adam G C, Likharev K K and Strukov D B 2015 Training and operation of an integrated neuromorphic network based on metal-oxide memristors *Nature* **521** 61–4

[47] Bennett C H, Parmar V, Calvet L E, Klein J-O, Suri M, Marinella M J and Querlioz D 2019 Contrasting advantages of learning with random weights and backpropagation in non-volatile memory neural networks *IEEE Access* **7** 1–16